\documentclass[12pt,preprint]{aastex}

\accepted{}

\slugcomment{Submitted to ApJ}

\newcommand{\ms}{\mbox{m s$^{-1}~$}}
\newcommand{\kms}{\mbox{km s$^{-1}~$}}
\newcommand{\msun}{M$_{\odot}$}

\newcommand{\mjup}{M$_{\rm JUP}$}

\newcommand{\msini}{$M \sin i~$}

\newcommand{\vsini}{V$sin i~$}

\newcommand{\simgt}{\lower.5ex\hbox{$\; \buildrel > \over \sim \;$}}
\newcommand{\simlt}{\lower.5ex\hbox{$\; \buildrel < \over \sim \;$}}

\shortauthors{Butler {\it et~al.\/}}
\shorttitle{Two New Planets}
\received{2004 June 10}
\begin{document}

\title{Multiple Companions to HD 154857 \& HD 160691$~^{1}$}

\author{Chris McCarthy\altaffilmark{2,3},
R. Paul Butler\altaffilmark{2},
C. G. Tinney\altaffilmark{4},
Hugh R. A. Jones\altaffilmark{5},
Geoffrey W. Marcy\altaffilmark{6,3},
Brad Carter\altaffilmark{7},
Alan J. Penny\altaffilmark{8},
Debra A. Fischer\altaffilmark{3,6}}

\authoremail{paul@dtm.ciw.edu}

\altaffiltext{1}{Based on observations obtained at the
Anglo--Australian Telescope, Siding Spring, Australia.}

\altaffiltext{2}{Department of Terrestrial Magnetism, Carnegie Institution
of Washington, 5241 Broad Branch Road NW, Washington DC, USA 20015-1305}

\altaffiltext{3}{Department of Physics and Astronomy,
San Francisco State University, San Francisco, CA, USA 94132}

\altaffiltext{4}{Anglo--Australian Observatory, P.O. Box 296,
Epping, NSW 1710, Australia}

\altaffiltext{5}{Astrophysics Research Institute, Liverpool John Moores
University, Twelve Quays House, Egerton Wharf, Birkenhead CH41 1LD, UK}

\altaffiltext{6}{Department of Astronomy, University of California,
Berkeley, CA USA 94720}

\altaffiltext{7}{Faculty of Sciences, University of Southern Queensland,
Toowoomba, Queensland 4350, Australia}

\altaffiltext{8}{Rutherford Appleton Laboratory,
Chilton, Didcot, Oxon, OX11 0QX, UK}

\begin{abstract}
Precise Doppler measurements from the AAT/UCLES spectrometer reveal
two companions to both HD 154857 and HD 160691.  The inner companion
to HD 154857 has a period of 398 d, an eccentricity of 0.51, and a
minimum mass of 1.8 \mjup.  An outer companion has a period much
longer than 2 years and is currently detected only as a linear trend
of 14 \ms per year.  The inner companion to HD 160691, previously
announced from AAT data, has a period of 645 d, an eccentricity of
0.20, and a minimum mass of 1.7 \mjup.  For the outer planet, whose
orbit is less well constrained, a two Keplerian fit
yields a period of 8.2 yr, an eccentricity of 0.57, and a minimum mass
of 3.1 \mjup.  With these orbital parameters, its maximum separation
from the star of 0.4 arcsec makes it a viable target for direct
imaging.

\end{abstract}


\section{Introduction}
\label{intro}

All $\sim$110 extrasolar planets\footnote{Count based on list
maintained at www.exoplanets.org as of 10 June 2004} discovered
within 50 parsecs have been found from precision Doppler surveys
over the past decade.  It is these nearest stars with planets
that are of the greatest interest for immediate follow up by 
HST astrometry (Benedict et al. 2002) and spectroscopy
(Charbonneau et al. 2002).  Next generation follow up missions
include precision astrometry with the Space Interferometry Mission
, and direct imaging of jovian and terrestrial planets
with giant ground--based telescopes
such as TMT\footnote{http://tmt.ucolick.org} (CELT),
GSMT
\footnote{http://klwww.aura-nio.noao.edu/science}
, Giant Magellan
\footnote{http://www.helios.lsa.astro.umich.edu/magellan/sciencecase.php}, 
and OWL
\footnote{http://www-astro.physics.ox.ac.uk/~imh/ELT},
and space--based telescopes such as TPF and DARWIN.


Precision Doppler surveys are most sensitive to planets orbiting
sun--like stars older than 2 Gyr that have no stellar companions
within 3 arc--seconds.  Stars of spectral type earlier than late F
rotate rapidly and have few absorption lines.  Stars younger than 2
Gyr have photospheric magnetic fields and spots, which cause an
intrinsic velocity ``jitter'' that correlates with chromospheric
activity (Wright et al. 2004).  Stars with close companions can not be
cleanly separated on a spectrometer entrance slit.  Coincidentally
single stars that are older than 2 Gyr represent the ideal targets for
habitable terrestrial planet searches, and by extension searches for
extraterrestrial intelligence (SETI) (Turnbull \& Tarter 2003).
There are roughly 2,000 FGKM stars within 50 parsecs that meet these
criteria.

Over the past 5 years the number of stars being monitored by
our precision Doppler programs has increased from $\sim$500
to nearly all 2,000 primary targets within 50 pc.  We
are carrying out northern hemisphere surveys with the Lick 3--m telescope
(Fischer et al. 2003) and Keck 10--m telescope (Vogt et al. 2002), and
southern surveys with the AAT 3.9--m (Tinney et al. 2003) and the
Magellan 6.5--m telescopes.  These programs have yielded 75 of the
110 published extrasolar planets (Butler et al. 2002).

In 2002 the guaranteed allocation to our AAT planet search
program was increased from 20 to 32 nights per year.  In
response two changes were made.  Sixty new stars were added,
and observations with S/N $>$200 were demanded for all stars.
In this paper we report two new planets from this survey, a
planet around HD 154857 including evidence of a second more
distant companion, and a second planet around HD 160691.

\section{Doppler Velocities and Keplerian Fits}

High resolution spectra, R $\sim$ 50000, are taken with the
UCLES echelle spectrometer (Diego et al. 1990) on the 3.9--m
Anglo--Australian Telescope.  These spectra span the wavelength
range from 4820--8550 \AA.  An iodine absorption cell
(Marcy \& Butler 1992) provides wavelength calibration from
5000 to 6100 \AA.  The spectrometer PSF is derived from the detailed
shapes of the embedded iodine lines (Valenti et al. 1995;
Butler et al. 1996).

Figure 1 shows four stable stars from the AAT.  These observations
span the time interval and the spectral types of the two planet
bearing stars reported in this paper.  The Doppler analysis
technique is outlined in Butler et al. (1996).  Over the past
year numerous improvements have been made to this algorithm.
The long term single--shot precision of chromospherically quiet
G and K stars is currently 2 \ms.  Similar systems on the
Lick 3--m, Keck 10--m and VLT2 8--m telescopes currently yield
photon--limited precision of 3 \ms (Butler \& Marcy 1997;
Vogt et al. 2000; Butler et al. 2001; Butler et al. 2004,
Marcy et al. 2004).

\subsection{HD 154857}

Houk \& Cowley (1975) classify HD 154857 (HIP 84069, SAO 244491) as a
G5 dwarf, consistent with its color, $B-V$ $=$ $0.65$.  The HIPPARCOS
 derived distance, 68.5 pc (ESA 1997; Perryman et al. 1996), yields an absolute V
magnitude, $M_{\rm V}$ = 3.07, $\sim$2 magnitudes above the main
sequence, suggesting that the star is evolved.  High metallicity alone is unlikely
to produce such a large displacement from the main sequence.  Like most evolved
stars moving toward subgiant status, HD 154857 is chromospherically
quiet with log R'HK = - 5.14 (Henry et al. 1996), but not in a Maunder
Minimum state (Wright 2004).  The
star is photometrically stable  with a scatter of 0.008 mag, consistent with
HIPPARCOS measurement error (ESA 1997).

Spectral synthesis matched to the spectrum of HD 154857
(Fischer \& Valenti 2004) yields Teff  $=$ $5628$ ($\pm$40) K,
log g $=$ $4.06$ ($\pm$0.05), and \vsini $=$ $3.31$ ($\pm$0.3) \kms,
consistent with a G5 dwarf evolving toward subgiant status.
The derived metallicity, [Fe/H] $=$ $-0.23$ ($\pm$0.03), is slightly lower
than for typical field stars (Reid 2001).  Interpolation on the grid of stellar models 
of Girardi et al. (2002) yields a stellar mass of 1.17 ($\pm$0.05) \msun \ and an
age of 5 Gyr.   

A total of 18 observations of HD 154857, taken between 2002 Apr and
2004 Feb, are shown in Figure 2 and listed in Table 1.  The median
value of internal measurement uncertainties is 3.39 \ms.  The dashed
line shows the best--fit single Keplerian model.  The RMS to this fit
is 7.66 \ms, yielding $\chi_{\nu}$ = 2.21.  The poor quality of this
fit motivated the addition of a linear trend to the Keplerian.  The
solid line shows the resulting best--fit Keplerian plus linear trend.
The RMS to this fit is 3.15 \ms, yielding $\chi_{\nu}$ = 0.87.  The
period of the Keplerian is 398.5 d, the semiamplitude is 52 \ms, and
the eccentricity is 0.51, yielding a minimum (\msini) mass of 1.80
\msun \ and a semimajor axis of 1.11 AU.  The linear trend of -14 \ms
per year is presumably due to an additional companion with a period
much longer than 2 years and a semiamplitude greater than 14 \ms.
Astrometry from HIPPARCOS reveals no wobble at the level of 1 mas (ESA
1997).  This places no useful constraint on the outer companion as its
orbital period is longer than the HIPPARCOS mission duration of 3
years, allowing any astrometric wobble to be absorbed in the solution
for proper motion.

\subsection{HD 160691}

The Bright Star Catalog (Hoffleit \& Jaschek 1982) assigns a spectral type
of G3 IV-V to HD 160691 (HR 6585, HIP 86796, GL 691, Mu Ara), in
reasonable agreement with numerous other determinations (see SIMBAD)
that typically yield G5 V.  The HIPPARCOS distance of HD 160691 
is 15.3 pc, yielding an absolute magnitude of $M_{\rm V}$ $=$ $4.23$.
The star is photometrically stable within HIPPARCOS measurement error,
exhibiting photometric scatter of 0.002 magnitudes, and is chromospherically
inactive, with log$R$'(HK) = $-$5.02 (Henry et al. 1996).

Spectral synthesis models fit to our high resolution spectrum of HD 160691
(Fischer \& Valenti 2004) yield Teff  $=$ $5807$ ($\pm$30) K,
log g $=$ $4.37$ ($\pm$0.05), and \vsini $=$ $4.07$ ($\pm$0.3) \kms.
The derived metallicity, [Fe/H] $=$ $+0.263$ ($\pm$0.03), is consistent
with previous determinations (Favata et al. 1997; Bensby et al. 2003).
Interpolation on the grid of stellar models by Girardi et al. (2002) yields a
stellar mass of 1.08 ($\pm$0.05) \msun \ and an age of 6 Gyr.

Based on 2 years of AAT data, Butler et al. (2001) reported a
planet with a 2 year period orbiting HD 160691.  With 4 years
of data, Jones et al. (2003) confirmed this planet and found an
additional linear trend.  Table 2 lists the 45 observations of
HD 160691 that now span 5.8 years.  These measurements, shown
in Figure 3, have a median internal uncertainty of 2.76 \ms.
The RMS to the best--fit Keplerian model plus a linear trend (solid line)
is 8.84 \ms.  With $\chi_{\nu}$ = 2.60 this model no longer
adequately fits the data.

Figure 4 shows a double Keplerian fit to the data.  The periods of
these two Keplerian models are 1.76 and 8.17 years, with orbital
eccentricities of 0.20 and 0.57, semiamplitudes of 38 and 51 \ms,
minimum (\msini) masses of 1.67 and 3.10 \mjup, and semimajor axes of
1.50 and 4.16 AU respectively.  This double Keplerian model represents
a significant improvement over the single Keplerian model with a
linear trend, with residuals yielding RMS = 4.66 \ms and $\chi_{\nu}$
= 1.60.  Nonetheless, other fits, with somewhat different periods and
eccentricities may have only slightly worse values of RMS scatter and
$\chi_{\nu}$.  Multi-Keplerian best fits can occasionally yield
solutions which are not dynamically stable (Gozdziewski, Konacki \&
Maciejewski, 2003).  Sophisticated dynamical analyses of this system,
such as those which impose the assumption of orbital stability, employ
Jacobi coordinates (e.g. Lee \& Peale, 2003; Gozdziewski, Konacki \&
Maciejewski, 2003), and which use genetic algorithms to derive superior
inital guesses are warrented, but beyond the scope of the present
paper.  Table 3 shows the orbital parameters for these two planets as
well as the those of the companion to HD 154857. 

Figure 5 shows each of the Keplerian orbits separately.  While the
period of the inner planet is tightly constrained with more than 3
full orbits observed, the outer planet is somewhat under constrained
with observations spanning only 70\% of the full period.  HIPPARCOS
astrometry reveals no wobble above standard errors of $\sim$1 mas.
The outer companion would cause a wobble easily absorbed into the
solution for proper motion.

Based on these fits, the maximum angular separation of the outer
planet, HD 160691c, from the star is expected to be $\alpha =
a(1+e)/d$ = 0.43 arcsec.  This planet is detectable by future imaging
systems that achieve resolution ($\sim 4 \lambda / d$) of less than
0.4 arcsec and that carefully suppresses the wings of the PSF due to
mirror roughness, including proposed space--based programs such as
ECLIPSE (Easley \& Trauger 2003), and ground based efforts at
VLT\footnote{http://www.eso.org/instruments} and
Gemini\footnote{http://www.gemini.edu/science}.

Future astrometry with precision of better than 1 mas, such as that
offered by the Hubble Space Telescope and its Fine Guidance Sensor
(Benedict et al. 2002) or by SIM, can also detect HD
160691c, however a duration of $\sim$ 8 yr would be required.

\section{Discussion}

Multiple planet systems are revealed considerably more slowly than
single--planets.  For ``hierarchical'' systems with widely separated
planets, the inner planet usually induces the largest Doppler wobble
in the host star, making it typically detectable first, as occurred
with 55 Cancri (Marcy et al. 2002) and $\upsilon$ And (Butler et al. 1999;
Butler et al. 1997).  For cases where the periods of both planets are
short, the most massive planet may be revealed first, as with GL 876
(Marcy et al. 2001).  Detectability stems from both the amplitude of
the Doppler signal and the number of cycles that have transpired during
the string of observations.

Both of the systems reported in this paper are of the first type, with
the inner planet detected first.  For the case of HD 154857, the
presumed outer companion is detected only as a linear trend at this
time.  Similarly, the existence of the second companions to HD 160691
and to 55 Cancri were first reported as a linear trends (e.g., Jones
et al. 2002).

Based on the results of Jones et al. (2002), Bois et al. (2003)
carried out a study of the dynamic stability of the HD 160691 system.
They concluded that the two planets must be in a 2:1 mean motion
resonance, and that the outer companion should be in a highly eccentric
(e $>$ 0.52) orbit.  With the identical data set, Gozdziewski et
al. (2003) conclude that this system cannot be in a 2:1 mean motion
resonance, and that if the semimajor axis of the outer companion is
smaller than 5.2 AU the system should be in a low-order mean motion
resonance (``3:1, 7:2, 4:1, 5:1, or to their vicinity''), and that the
eccentricity of the outer companion should be less than 0.5.

With data covering only $\sim$70\% of the orbit, the orbital
parameters of the outer companion to HD 160691 have relatively large
uncertainties.  Nonetheless the lower limit of the period of HD
160691c is 8 yr, ruling out a 2:1 mean motion resonance with the
inner planet ($P$=1.76 yr).  Similarly, mean--motion resonances of 3:1
and 4:1 seem ruled out.  A 5:1 mean motion resonance remains plausible
as is one of 9:2 ratio.  The orbital eccentricity of the outer planet,
however, remains poorly constrained.

Both of these systems highlight the importance of long term monitoring
and of Doppler measurement precision.  Neither the second planet to HD
160691 nor the linear trend to HD 154857 would be detectable with
measurement uncertainties greater than 7 \ms.   The outer planet around
HD 160691 is one of only a few known extrasolar planets that is separated
from its host star at apastron by over 0.4 arcsec, making it a prime target for
future efforts designed to directly image extrasolar planets.

\acknowledgements

We gratefully acknowledge the superb technical support at the
Anglo--Australian Telescope which has been critical to the success of
this project -- in particular E. Penny, R. Paterson, D. Stafford,
F. Freeman, S. Lee, J. Pogson, and G. Schaffer.  We further
acknowledge the UK and Australian government support of the
Anglo-Australian Telescope through their PPARC and DETYA funding to
HRAJ, AJP, \& CGT. We received support from NSF grant AST-9988087 and
travel support from the Carnegie Institution of Washington (to RPB),
NASA grant NAG5-8299 and NSF grant AST95-20443 (to GWM), and from Sun
Microsystems.  We thank the Australian and UK Telescope assignment
committees (ATAC \& PATT) for allocations of telescope time.  This
research has made use of NASA's Astrophysics Data System, and the
SIMBAD database, operated at CDS, Strasbourg, France.

\clearpage
\vspace{-0.5in}
\begin{figure}
\epsscale{0.8}
\plotone{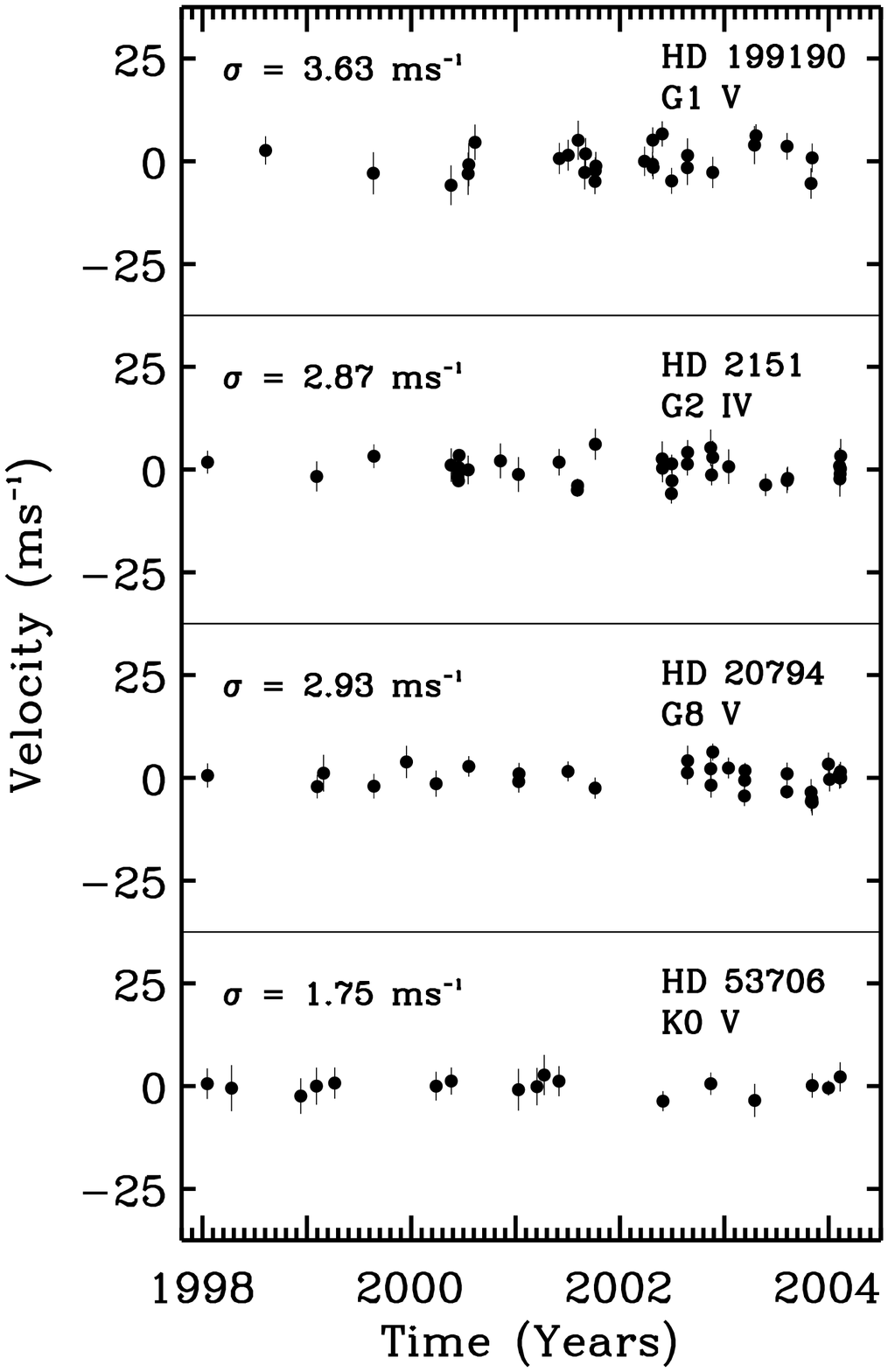}
\caption{AAT Doppler velocities of chromospherically quiet
G and K dwarfs and subgiants.  These observations span the
full 6 years of the AAT Planet Search Project.}
\label{fig1}
\end{figure}

\begin{figure}
\vspace{-0.5in}
\plotone{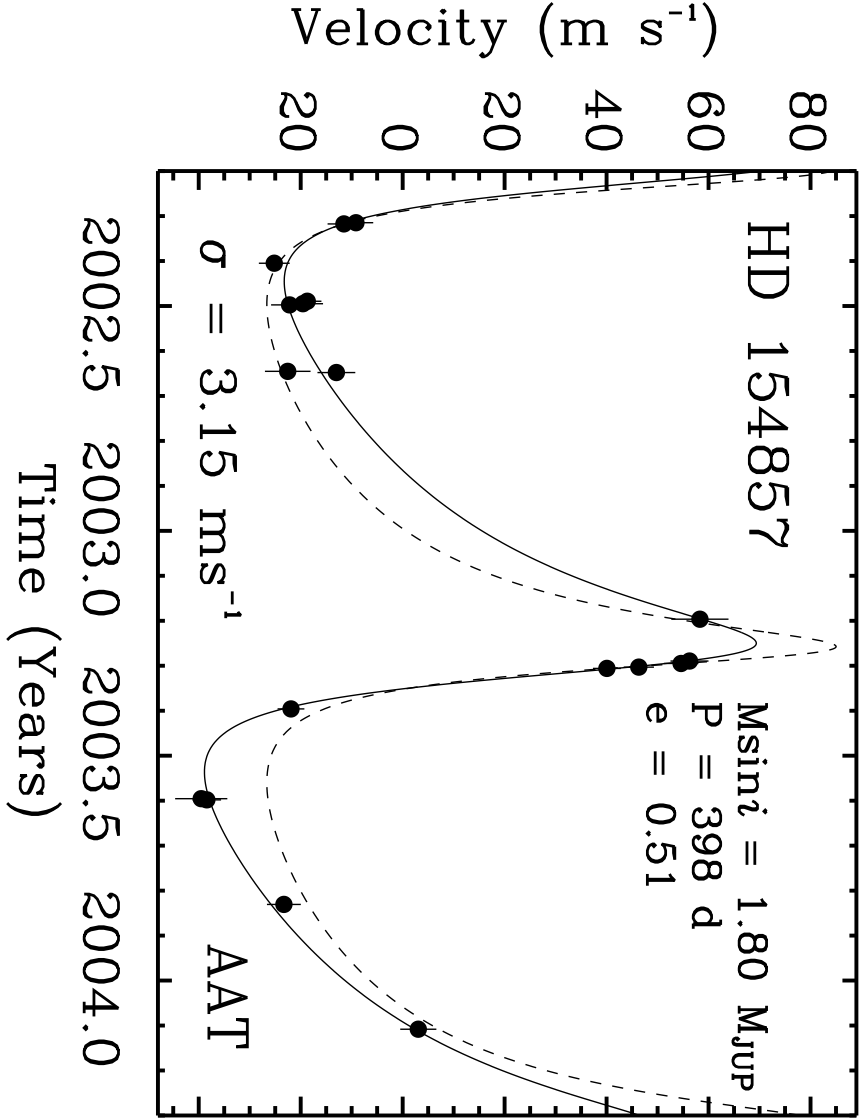}
\caption{Doppler velocities for HD 154857.  The dashed line is the
best--fit Keplerian.  The RMS of 7.66 \ms is significantly worse
than the median measurement uncertainty of 3.39 \ms.  The solid
line shows the best--fit Keplerian model with an additional linear trend.
The RMS of 3.15 \ms is consistent with measurement uncertainty.
The period of this Keplerian model is 398.5 d, the semiamplitude is
52 \ms, the eccentricity is 0.51, yielding a minimum mass of
1.80 \mjup \ and a semimajor axis of 1.11 AU.  The linear trend
is presumably due to a companion with a period much longer than
2 years and a mass significantly greater than 1 \mjup.}
\label{fig2}
\end{figure}

\begin{figure}
\vspace{-0.5in}
\plotone{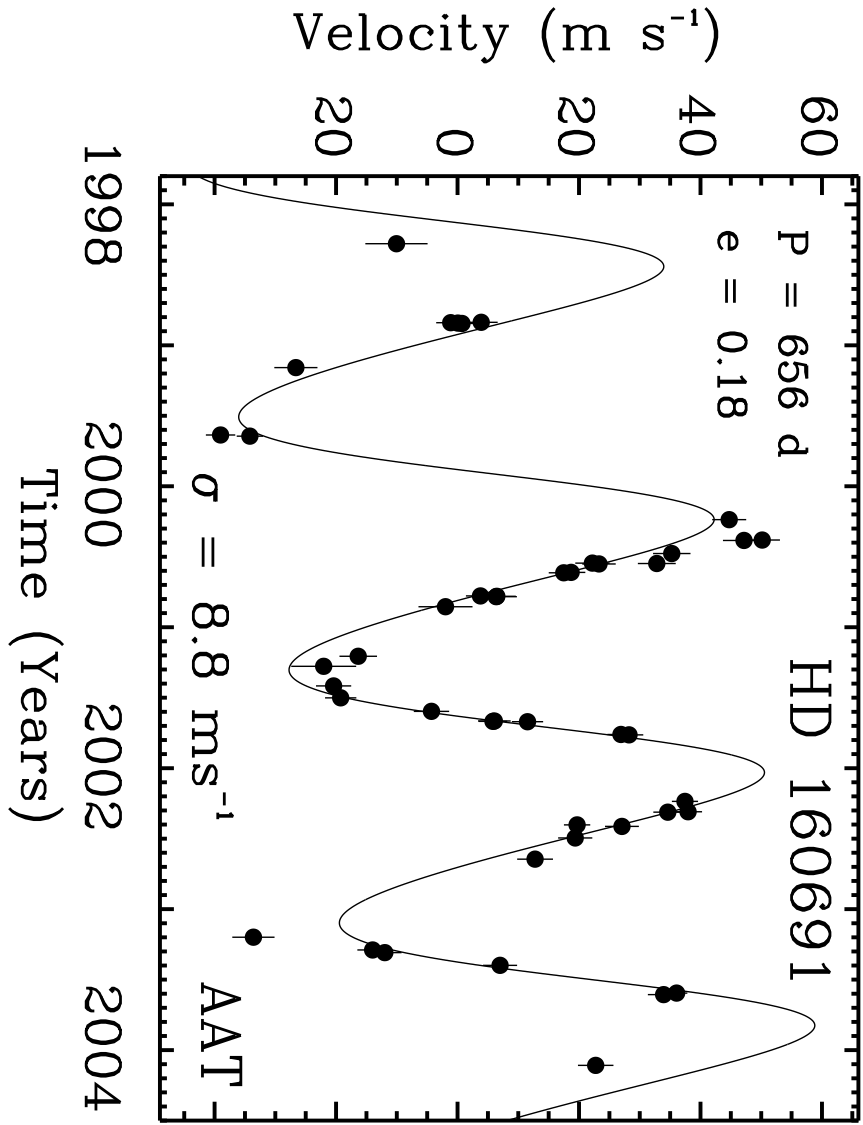}
\caption{Doppler velocities for HD 160691.  The solid line is
the best--fit Keplerian plus linear trend.  The RMS to this
model is 8.84 \ms, significantly worse than the median internal
measurement uncertainty of 2.76 \ms.}
\label{fig3}
\end{figure}

\begin{figure}
\vspace{-0.5in}
\plotone{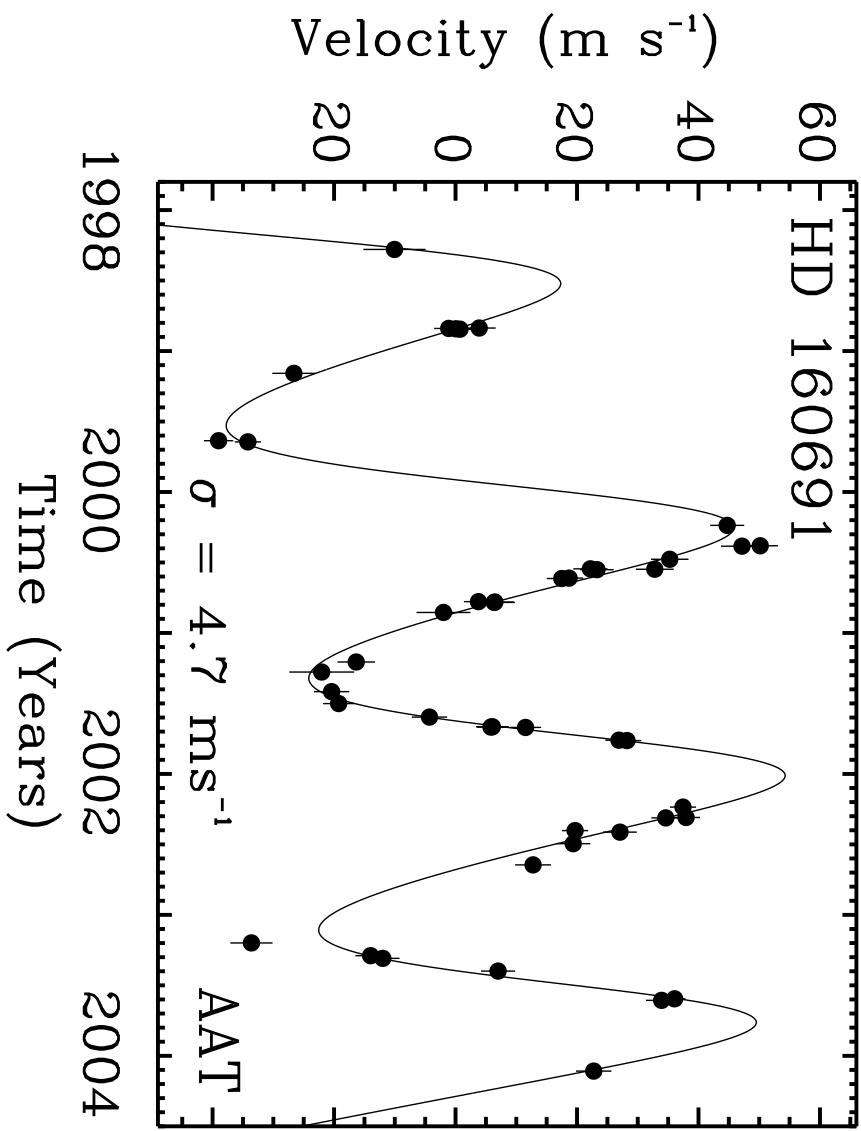}
\caption{A double Keplerian fit to HD 160691.  The inner Keplerian
has a period of 645 d, and eccentricity of 0.20, and a semiamplitude
of 38 \ms, yielding a minimum mass of 1.68 \mjup \ and a semimajor
axis of 1.50 AU.  The outer Keplerian has a period of 8.2 yr,
an eccentricity of 0.57, and a semiamplitude of 51 \ms, yielding
a minimum mass of 3.1 \mjup \ and a semimajor axis of 4.16 AU.
The RMS to this model is 4.66 \ms  (1.69 $\sigma$)}
\label{fig4}
\end{figure}

\begin{figure}
\vspace{-0.5in}
\plotone{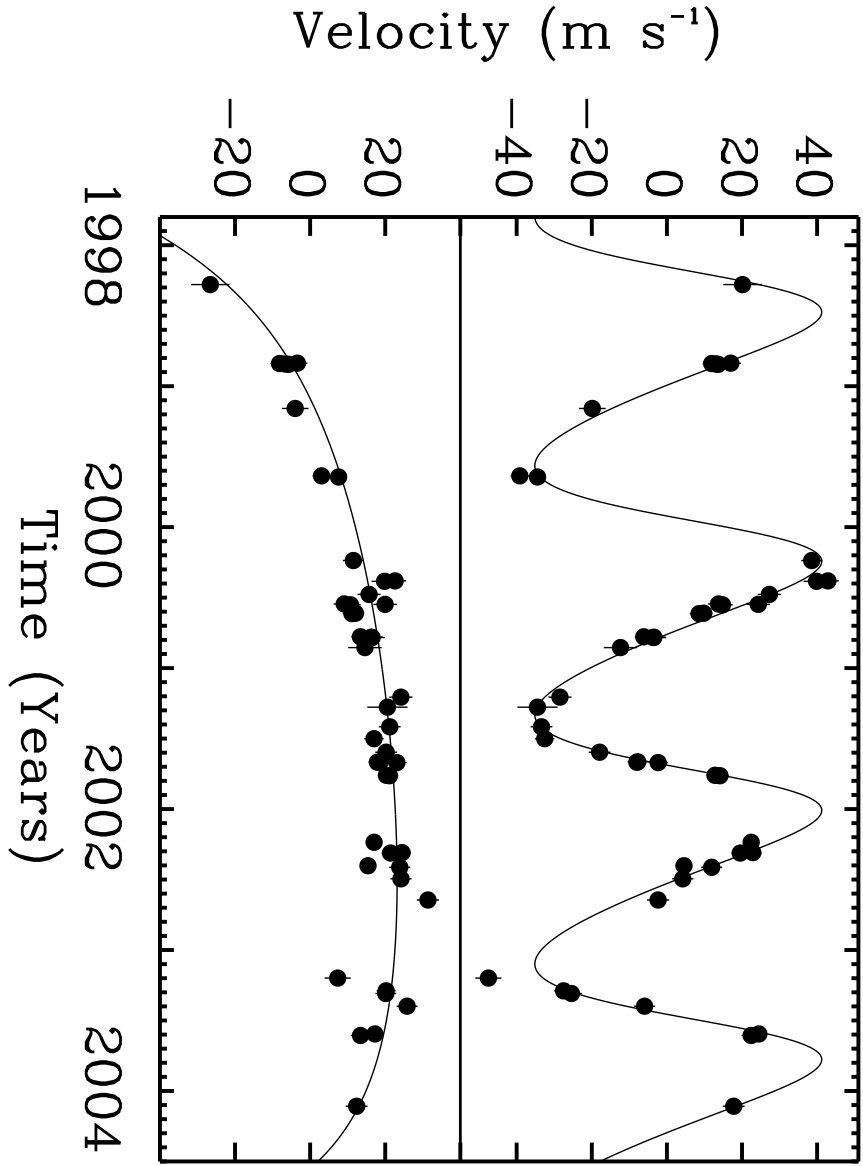}
\caption{Individual Keplerian models to HD 160691.  Each of the
two Keplerian models from Figure 4 is shown separately.  The
top figure shows the 645 d Keplerian, while the bottom
figure shows the 8.17 yr Keplerian.  Approximately 70\%
of the outer Keplerian has been observed.}
\label{fig5}
\end{figure}

\clearpage

\begin{deluxetable}{rrr}
\tablenum{1}
\tablecaption{Velocities for HD 154857}
\label{velbd-103166}
\tablewidth{0pt}
\tablehead{
JD & RV & error \\
(-2450000)   &  (m s$^{-1}$) & (m s$^{-1}$)
}
\startdata
\tableline
  2389.2358  &   -14.6  &  3.4 \\
  2390.2122  &   -16.9  &  3.2 \\
  2422.1371  &   -30.6  &  3.0 \\
  2453.0198  &   -24.1  &  2.8 \\
  2455.0253  &   -25.0  &  3.9 \\
  2455.9766  &   -27.6  &  3.6 \\
  2509.9485  &   -28.0  &  4.5 \\
  2510.9162  &   -18.4  &  3.7 \\
  2711.2461  &    52.9  &  5.6 \\
  2745.2427  &    50.8  &  3.7 \\
  2747.2115  &    49.2  &  3.4 \\
  2750.1777  &    40.9  &  3.0 \\
  2751.2291  &    34.6  &  2.7 \\
  2784.1264  &   -27.3  &  2.6 \\
  2857.0297  &   -44.9  &  5.1 \\
  2857.9860  &   -43.8  &  2.8 \\
  2942.9120  &   -28.7  &  3.3 \\
  3044.2691  &    -2.4  &  3.5 \\
\enddata
\end{deluxetable}

\clearpage

\begin{deluxetable}{rrr}
\tablenum{2}
\tablecaption{Velocities for HD 160691}
\label{vel160691}
\tablewidth{0pt}
\tablehead{
JD & RV & error \\
(-2450000)   &  (m s$^{-1}$) & (m s$^{-1}$)
}
\startdata
\tableline
   915.2911  &   -18.8  &  5.1 \\
  1118.8874  &    -4.9  &  2.7 \\
  1119.9022  &    -9.9  &  2.4 \\
  1120.8870  &    -8.7  &  2.4 \\
  1121.8928  &    -8.0  &  2.6 \\
  1236.2864  &   -35.4  &  3.6 \\
  1410.8977  &   -47.8  &  2.4 \\
  1413.8981  &   -43.0  &  2.2 \\
  1630.3042  &    36.0  &  2.8 \\
  1683.0926  &    41.4  &  2.9 \\
  1684.1320  &    38.4  &  3.5 \\
  1718.1184  &    26.5  &  3.1 \\
  1742.9096  &    13.4  &  2.8 \\
  1743.9240  &    24.0  &  3.1 \\
  1745.0440  &    14.5  &  2.8 \\
  1766.9330  &     9.9  &  2.3 \\
  1767.9689  &     8.7  &  2.5 \\
  1827.8973  &    -5.0  &  2.4 \\
  1828.8866  &    -2.3  &  3.1 \\
  1829.8890  &    -2.4  &  3.4 \\
  1855.9058  &   -10.7  &  4.4 \\
  1984.2618  &   -25.1  &  3.1 \\
  2010.2829  &   -30.8  &  5.3 \\
  2061.1132  &   -29.2  &  2.9 \\
  2091.9807  &   -28.0  &  2.6 \\
  2126.9766  &   -13.1  &  2.9 \\
  2151.9693  &    -2.7  &  2.7 \\
  2152.9493  &    -2.9  &  2.3 \\
  2153.8626  &     2.8  &  2.6 \\
  2186.9095  &    18.1  &  2.2 \\
  2187.8879  &    19.5  &  2.4 \\
  2360.3245  &    28.7  &  2.2 \\
  2387.1722  &    29.2  &  2.3 \\
  2388.2097  &    25.9  &  2.4 \\
  2421.1696  &    10.9  &  2.2 \\
  2425.1226  &    18.3  &  2.8 \\
  2455.0437  &    10.6  &  2.8 \\
  2509.9587  &     4.0  &  3.0 \\
  2712.1925  &   -42.4  &  3.5 \\
  2745.2553  &   -22.7  &  2.5 \\
  2752.1799  &   -20.7  &  2.8 \\
  2785.1422  &    -1.8  &  2.8 \\
  2857.0201  &    27.3  &  1.8 \\
  2860.9744  &    25.2  &  2.6 \\
  3044.2886  &    14.0  &  2.9 \\
\enddata
\end{deluxetable}

\clearpage

\begin{deluxetable}{lccc}
\tablenum{3}
\tablecaption{Orbital Parameters}
\label{orbit}
\tablewidth{0pt}
\tablehead{
\colhead{Parameter} & \colhead{HD 154857\tablenotemark{a}} & \colhead{HD 160691b} & \colhead{HD 160691c}
}
\startdata
Orbital period $P$ (d) &  398.5 (9) & 645.5 (3)     & 2986 (30) \\
Velocity amp. $K$ (m\,s$^{-1}$)     & 52 (5)        & 38 (2) & 51 (7) \\
Eccentricity $e$  & 0.51 (0.06)     & 0.20 (0.03)   & 0.57 (0.1) \\
$\omega$ (deg)    & 50 (11)         & 294 (9)       & 161 (8) \\
Periastron Time (JD) & 2451963 (10) & 2450645.5 (4) & 2450541 (96) \\
Msini (\mjup)     &  1.80 (0.27)    & 1.67  (0.11)  & 3.10 (0.71) \\
a (AU)            &  1.11 (0.02)    & 1.50  (0.02)  & 4.17 (0.07) \\
RMS (\ms)         &  3.15           & 4.66          &  --- \\
\enddata
\tablenotetext{a}{Additional Slope is -14.3 $\pm$3.5 \ms per yr.}

\tablecomments{Errors in fit parameters (1-$\sigma$, shown in
parenthesis), are derived from formal Monte Carlo tests.  Error in the
derived quantities $Msini$ and $a$ reflect the quadrature sum of
1-$\sigma$ errors in fit Keplerian parameters and stellar mass.}

\end{deluxetable}

\clearpage

\clearpage

\begin{thebibliography}{}

\bibitem[Benedict { et~al.} 2002]{Benedict03}
Benedict, G.~F., McArthur, B.~E., Forveille, T.,
Delfosse, X., Nelan, E., Butler, R.~P., Spiesman, W.,
Marcy, G.~W., Goldman, B., Perrier, C.,
Jefferys, W.~H. \& Mayor, M. 2002,
\newblock { ApJ Letters, }, 581, L115.

\bibitem[Bensby { et~al.} 2003]{Bensby03}
Bensby, T., Feltzing, S., \& Lundstroem, I. 2003,
\newblock { A\&A, }, 410, 527.

\bibitem[Bois { et~al.} 2003]{Bois03}
Bois, E., Kiseleva-Eggleton, L.,
Rambaux, N., \& Pilat-Lohinger, E. 2003
\newblock { ApJ, }, 598, 1392.

\bibitem[Butler { et~al.} 1996]{BuMaWi96}
Butler, R.~P., Marcy, G.~W., Williams, E., McCarthy, C.,
Dosanjh, P., \& Vogt, S.~S.  1996,
\newblock { PASP }, {108}, 500

\bibitem[Butler \& Marcy 1997]{BuMa97}
Butler, R.~P. \& Marcy, G.~W. 1997, in ``Brown Dwarfs and
Extrasolar Planets'', ed. R.Rebolo, E.L.Martin,
\& M.R.Zapatero Osorio, ASP Conference Series 134, p. 162. 

\bibitem[Butler { et~al.} 1997]{BuMaWi97}
Butler, R.~P., Marcy, G.~W., Williams, E.,
Hauser, H., \& Shirts, P. 1997,
\newblock { ApJ Letters }, 474, L115.

\bibitem[Butler { et~al.} 1999]{BuMaWi99}
Butler, R.~P., Marcy, G.~W., Fischer, D.~A.,
Brown, T.~M., Contos, A.~R., Korzennik, S.~G.,
Nisenson, P., \& Noyes, R.~W. 1999,
\newblock { ApJ }, 526, 916.

\bibitem[Butler { et~al.} 2001]{AAT01}
Butler, R.~P., Tinney, C.~G., Marcy, G.~W.,
Jones, H.~R.~A., Penny, A.~J. \& Apps, K. 2001, 
\newblock { ApJ }, 555, 410.

\bibitem[Butler { et~al.} 2002]{HD83443:02}
Butler, R.~P., Marcy, G.~W., Vogt, S.~S., Tinney, C.~G., 
Jones, H.~R.~A., McCarthy, C., Penny, A.~J.,
Apps, K., \& Carter, B.~D.  2002, 
\newblock { ApJ }, 578, 565.

\bibitem[Butler { et~al.} 2004]{VLT:04}
Butler, R.~P., Bedding, T.~R., Kjeldsen, H., McCarthy, C.,
O'Toole, S.~J., Tinney, C.~G., Marcy, G.~W., Wright, J.~T. 2004,
\newblock { ApJ Letters}, 600, L75.

\bibitem[Charbonneau {  et~al.} 2002]{Charbon02}
Charbonneau, D, Brown, T.~M., Noyes, R.~W. \& Gilliland, R.~L. 2002,
\newblock {  ApJ, } 568, 377. 

\bibitem[Diego {  et~al.} 1991]{diego:90}
Diego, F., Charalambous, A., Fish, A.~C., \& Walker, D.~D.  1990, 
Proc. Soc. Photo-Opt. Instr. Eng., 1235, 562

\bibitem[ESA 1997]{esa97}
ESA 1997, Vizier Online Data Catalog, HIPPARCOS, 1239.

\bibitem[Easley \& Trauger 2004]{easley04}
Easley, M.~A. \& Trauger,J.T. 2004, 
in "UV/Optical/IR space Telescopes: Innovative Technologies and Concepts",
ed MacEwen,H.A., in Proceedings of the SPIE, v.5166, p172.

\bibitem[Favata {  et~al.} 1997]{Fav97}
Favata, F., Micela, G., \&  Sciortino, S. 1997,
\newblock {  A\&A }, {323}, 809

\bibitem[Fischer {  et~al.} 2003]{Fischer2003}
Fischer, D.~A., Marcy, G.~W., Butler, R.~P.,
Vogt, S.~S., Henry, G.~W., Pourbaix, D.,
Walp, B., Misch, A.~A. \& Wright, J.~T. 2003,
\newblock {  ApJ, } 586, 1394. 

\bibitem[Fischer \& Valenti 2004]{FischVal04}
Fischer, D.~A. \& Valenti, J 2004,
\newblock {  ApJ, } submitted. 

\bibitem[Girardi {  et~al.} 2002]{Girardi02}
Girardi, L, Bertelli, G., Bressan, A., Chiosi, C.,
Groenewegen, M.~A.~T., Marigo, P.,
Salasnich, B. \& Weiss, A. 2002,
\newblock {  A\&A, } 391, 195. 

\bibitem[Gozdziewski { et~al.} 2003]{Goz03}
Gozdziewski, K., Konacki, M., \& Maciejewski, A.~J. 2003,
\newblock { ApJ }, 594, 1019

\bibitem[Henry {  et~al.} 1996]{Henry96}
Henry, T. J., Soderblom, D. R., Donahue, R. A., \& Baliunas, S. L. 1996,
\newblock { AJ, } {111}, 439

\bibitem[Hoffleit 1982]{Hoff82}
Hoffleit, D. \& Jaschek, C. 1982, {\it Yale Bright Star Catalogue},
4th edition (Yale University, New Haven).

\bibitem[Houk \& Cowley 1975]{Houk75}
Houk, N. \& Cowley, A.~P. 1975,
{\it Michigan Catalog of Two Dimensional Spectral Types for the HD Stars},
(Ann Arbor, University of Michigan).

\bibitem[Jones 2002]{jones02}
Jones, H.~R.~A., Butler, R.~P., Marcy, G.~W., Tinney, C.~G.,
Penny, A.~J., McCarthy, C., \& Carter, B.~D. 2002,
\newblock {  MNRAS }, 337, 1170.

\bibitem[Marcy \& Butler 1992]{MaBu92}
Marcy, G.~W. \& Butler, R.~P. 1992,
\newblock {  PASP }, {104}, 270.

\bibitem[Marcy {  et~al.} 2001]{gl876:01}
Marcy, G.~W., Butler, R.~P., Fischer, D.~A.,
Vogt, S.~S., Lissauer, J.~J., Rivera, E.~J. 2001,
\newblock { ApJ, } 556, 296. 

\bibitem[Marcy {  et~al.} 2002]{55can:02}
Marcy, G.~W., Butler, R.~P., Fischer, D.~A.,
Laughlin, G., Vogt, S.~S., Henry, G.~W., \& Pourbaix, D. 2002,
\newblock { ApJ, } 581, 1375.

\bibitem[Marcy {  et~al.} 2004]{marcy:04}
Marcy, G.~W., Butler, R.~P., Vogt, S.~S.,  Fischer, D.~A.,
Henry, G.~W., Laughlin, G., Wright, J.~T., \& Johnson, J. 2004, 
\newblock { ApJ, } submitted.

\bibitem[Perryman {  et~al.} 1997]{Perry97}
Perryman, M.~A.~C., et al. 1997, { A\&A }, 323, L49.
The HIPPARCOS Catalog

\bibitem[Reid 2002]{Reid02}
Reid, I.~N. 2002, PASP, 114, 306.

\bibitem[Tinney {  et~al.} 2003]{TinBut03}
Tinney, C.~G., Butler, R.~P., Marcy, G.~W., Jones, H.~R.~A.,
Penny, A.~J., McCarthy, C., Carter, B.~D., \& Bond, J. 2003,
\newblock {  ApJ, } 587, 423. 

\bibitem[Turnbull \& Tarter 2003]{TurnTart03}
Turnbull, M.~C. \& Tarter, J.~C. 2003,
\newblock {  ApJS, } 145, 181.

\bibitem[Valenti { et~al.} 1995]{VaBuMa95}
Valenti, J., Butler, R.~P., \& Marcy, G.~W. 1995,
\newblock { PASP }, {107}, 966

\bibitem[Vogt {  et~al.} 2000]{vogt:00}
Vogt, S.~S., Marcy,G.W., Butler,R.P., \& Apps,K. 2000,
\newblock { ApJ, } {536}, 902

\bibitem[Vogt {  et~al.} 2002]{vogt:02}
Vogt, S.~S., Butler, R.~P., Marcy, G.~W., Fischer, D.~A.,
Dimitri, P., Apps, K., Laughlin, G. 2002,
\newblock { ApJ, } {568}, 352.


\bibitem[Wright {  et~al.} 2004]{wrighta:04}
Wright, J.~T., Marcy, G.~W., Butler, R.~P. \& Vogt, S.~S. 2004,
\newblock { ApJ Supp, } in press.

\bibitem[Wright 2004]{wrightb:04}
Wright, J.~T. 2004,
\newblock { ApJ, } in press.

\end{thebibliography}
\end{document}